\documentclass[pdftex,twocolumn,epjc3]{svjour3}          

\RequirePackage[T1]{fontenc}

\smartqed  

\RequirePackage{graphicx}
\RequirePackage{mathptmx}      
\RequirePackage{flushend}
\RequirePackage[numbers,sort&compress]{natbib}
\RequirePackage[colorlinks,citecolor=blue,urlcolor=blue,linkcolor=blue]{hyperref}

\journalname{Eur. Phys. J. C}
\usepackage{graphicx}
\usepackage{amssymb}
\usepackage{amsmath}
\usepackage{color}
\usepackage{hyperref}
 \usepackage{tabu}

\def\barray{\begin{array}}
\def\earray{\end{array}}
\def\be{\begin{equation}}
\def\ee{\end{equation}}
\def\ben{\begin{equation} \nonumber}
\def\een{\end{equation}}
\def\ban{\begin{eqnarray*}}
\def\ean{\end{eqnarray*}}
\def\ba{\begin{eqnarray}}
\def\ea{\end{eqnarray}}

\def\({\left(}
\def\){\right)}

\graphicspath{{./fig/}}

\begin{document}

\title{Tachyon warm-intermediate inflation in the light of Planck data}
\author{Vahid  Kamali\thanksref{e,addr},
Spyros Basilakos\thanksref{e1,addr1}
        \and
        Ahmad  Mehrabi\thanksref{e2,addr} 
}
\thankstext{e}{e-mail: vkamali@basu.ac.ir}
\thankstext{e1}{e-mail:svasil@academyofathens.gr}
\thankstext{e2}{e-mail: Mehrabi@basu.ac.ir}
\institute{Department of Physics, Bu-Ali Sina University, Hamedan
65178, 016016, Iran\label{addr}
          \and
          Academy of Athens, Research Center for Astronomy \& Applied 
  Mathematics, Soranou Efessiou 4, 11-527, Athens, Greece\label{addr1}
}

\date{Received: date / Accepted: date}

\maketitle

\begin{abstract}
We study the main properties of the warm tachyon inflation model in the framework of RSII braneworld  based on Barrow's solution for the scale factor of the universe.
Within this framework we calculate analytically the basic slow roll parameters 
for different versions of warm inflation. 
We test the performance of this inflationary scenario against 
the latest observational data and we verify that the predicted 
spectral index and the tensor-to-scalar fluctuation ratio 
are in excellent agreement with those of {\it Planck 2015}.
Finally, we find that the current predictions are consistent 
with those of viable inflationary models.

\end{abstract}

\maketitle

\section{Introduction}
Standard inflation driven by an inflaton field traces back to early
efforts to alleviate the basic problems of the Big-Bang cosmology, namely
horizon, flatness and  
monopoles \cite{Guth:1980zm, Albrecht:1982wi}.
The nominal inflationary paradigm contains the slow-roll and the 
(P)reheating regimes.
In the slow-roll phase the kinetic energy (which has the canonical form here) 
of the scalar field is negligible with respect to the potential energy $V(\phi)$
which implies a deSitter expansion of the Universe.  
However, after the slow-roll epoch the kinetic energy 
becomes comparable to the potential energy and thus  
the inflaton field oscillates around the 
minimum and progressively the universe is filled by 
radiation \cite{Shtanov:1994ce,Kofman:1997yn}.

Nevertheless, other theoretical patterns suggested 
a possible way to treat the physics of the early universe. 
For example, in the so-called warm inflationary 
scenario the radiation production
occurs during the slow-roll epoch and the
reheating period is avoided \cite{Berera:1995ie, Berera:1996fm}.
The nature of the warm inflationary scenario 
is different with respect to that of the 
standard cold inflation. 
Warm inflation satisfies 
the condition $T>H$, where $T$ is the temperature 
and $H$ is the Hubble parameter, which implies that
the fluctuations of the inflaton field are thermal instead of quantum.
An obvious consequence of the above inequality is that in the case of warm inflation 
density perturbations arise from thermal 
fluctuations rather than quantum fluctuations
\cite{Hall:2003zp, Moss:1985wn, Berera:1999ws}.  
Specifically, thermal fluctuations are
produced during the warm inflationary epoch and they play 
a central role toward describing the CMB anisotropies and thus 
providing the initial seeds for the formation of large scale structures.  
Of course, after this epoch the universe enters in the radiation 
dominated phase as it should \cite{Berera:1995ie, Berera:1996fm}.
In order to achieve warm inflation one may use a tachyon scalar field
for which the kinetic term does not follow the canonical 
form (k-inflation \cite{ArmendarizPicon:1999rj}).
It has been found that tachyon fields which are associated with
unstable D-branes \cite{Sen:2002nu} can be responsible 
for the cosmic acceleration in early times 
\cite{Sen:2002an, Sami:2002fs, ArmendarizPicon:1999rj}. 

Notice, that tachyon potentials have the following two properties:
the maximum of the potential occurs when
$\phi\rightarrow 0$ while the corresponding minimum takes place
when $\phi\rightarrow \infty$. 
From the dynamical viewpoint one may obtain the equations of motion
using a special Lagrangian 
\cite{Gibbons:2002md} which is non-minimally coupled to gravity:
\begin{equation}\label{Lag}
L=\sqrt{-g}\left[\frac{R}{16\pi G}-V(\phi)\sqrt{1-g^{\mu\nu}\partial_{\mu}\phi\partial_{\nu}\phi}\right]\;.
\end{equation}
Considering a spatially flat  
Friedmann-Robertson-Walker (hereafter FRW) space-time
the stress-energy tensor is given by 
\begin{equation}\label{1.1}
T^{\mu}_{\nu}=\frac{\partial
L}{\partial(\partial_{\mu}\phi)}\partial_{\nu}\phi-g^{\mu}_{\nu}L={\rm diag}(-\rho_{\phi},p_{\phi},p_{\phi},p_{\phi})
\end{equation}{equation}
where $\rho_{\phi}$ and $p_{\phi}$
are the energy density and pressure of the scalar field. 
Combining the 
above set of equations one can derive 
\begin{equation}\label{1.2}
\rho_{\phi}=\frac{V(\phi)}{\sqrt{1-\dot{\phi}^2}}
\end{equation}
and
\begin{equation}\label{1.22}
P_{\phi}=-V(\phi)\sqrt{1-\dot{\phi}^2}
\end{equation}
Where $\phi$ is tachyon scalar field in unite of inverse Planck mass $M_{pl}^{-1}$, and $V(\phi)$ is potential associated with the tachyon field. 
In the past few years, there was an intense debate among 
cosmologists and particle physicists regarding those 
phenomenological models which can be produced in extra dimensions.   
For example, the reduction of higher-dimensional 
gravitational scale, down to
TeV-scale, could be presented by an 
extra dimensional scenario \cite{ArkaniHamed:1998rs,ArkaniHamed:1998nn,Antoniadis:1998ig}. In these scenarios, gravity field
propagates in the bulk while 
standard models of particles are confined to the lower-dimensional brane.
In this framework, the extra dimension induces additional terms 
in the first Friedmann equation \cite{Binetruy:1999ut,Binetruy:1999hy,Shiromizu:1999wj}.
Especially, if we consider a quadratic term in the energy density 
then we can extract an accelerated expansion of the early universe \cite{Maartens:1999hf,Cline:1999ts,Csaki:1999jh,Ida:1999ui,Mohapatra:2000cm}.
 In the current study we consider the 
tachyon warm inflation model in the framework of 
Randall-Sundrum II braneworld which contains
a single, positive tension brane and a non-compact extra dimension.

Following the lines of Ref.\cite{Herrera:2015aja}, we attempt
to study the main properties of the warm inflation  
in which the scale factor evolves as $a(t)\propto \exp(At^f)$, where $0<f<1$
("intermediate inflation"). 
In this case cosmic expansion evolves faster than the power-law inflation
($a \propto t^{p}$, $p>1$)
and slower than the standard deSitter one, $a(t)\propto {\rm exp}(H_{I}t)$
[$H(t)=H_{I}=$const.]. More details regarding the cosmic expansion  
in various inflationary solutions can be found in the paper of 
Barrow \cite{Barrow:1996bd}. 

In the current work, we investigate the possibility 
of using the intermediate solution in the case of 
warm tachyon inflation. Specifically, the structure of 
the article is as follows: In section II we briefly
discuss the main properties of the 
warm inflation, while in section III we provide the slow roll parameters.
In section IV we study the
performance of our predictions 
against the \textit{Planck 2015} data.
Finally, the main conclusions are
presented in section VI.



\section{Tachyon warm inflation}\label{setup}
Let us assume a flat, homogeneous and isotropic
Friedmann-Robertson-Walker (FRW) universe, 
in which the radiation era is endowed with the scalar field 
described by the Lagrangian (\ref{Lag})
in the context 
of the Randall-Sundrum II (RSII)  
brane \cite{Randall:1999vf}
. 
Following the notations of 
\cite{Shiromizu:1999wj, Binetruy:1999ut, Binetruy:1999hy,Berera:1995ie, Berera:1996fm, Herrera:2015aja} 
one may check that
the basic cosmological equations 
are
\begin{equation}\label{2.3}
H^2=\frac{8\pi}{3M_{pl}^2}\rho(1+\frac{\rho}{2\lambda})
\end{equation}{equation}
\begin{equation}\label{denn}
{\dot \rho}_{\phi}+3H(\rho_{\phi}+p_{\phi})=-\Gamma{\dot \phi}^2
\end{equation}
\begin{equation}\label{denn1}
{\dot \rho}_{\gamma}+3H(\rho_{\gamma}+p_{\gamma})=\Gamma {\dot \phi}^2
\end{equation}
where $\Gamma$ is the dissipation coefficient, in unit of $M_{pl}^5$. 
The latter two equations (\ref{denn}),(\ref{denn1}) imply 
the continuity equation, namely ${\dot \rho}+3H(\rho+p)=0$. 
 Notice, that Eqs.(\ref{denn}) and (\ref{denn1})
have been proposed by various authors such as 
\cite{Cai:2010wt,Herrera:2006ck,delCampo:2008fc,Setare:2012fg,Setare:2013dd,Zhang:2013waa}. In these studies the quantity $\Gamma \dot{\phi}^2$ is the dissipation term 
which is introduced phenomenologically in order to describe the 
nearly-thermal radiation bath that is the outcome of the warm 
inflationary scenario.
 It is well known that 
Tachyon inflation in its standard picture (cold inflation) 
suffers from a serious problem. In particular, 
reheating and matter creation are both problematic because the tachyon fields 
in such theories do not oscillate around the minimum of the potential 
\cite{Kofman:2002rh}. 
This problem can be alleviated in the context of warm inflation. 
In this scenario radiation production occurs during the slow-roll 
era which implies that reheating is avoided and thus the universe 
heats up and finally it enters in the radiation era \cite{Berera:1995ie} (See  Eqs.(\ref{denn}) and (\ref{denn1})).

In the above set of equations, an over-dot denotes
derivative with respect to time,
$\rho=\rho_{\phi}+\rho_{\gamma}$ and $p=p_{\phi}+p_{\gamma}$
($p_{\gamma}=\rho_{\gamma}/3$) are the total density and pressure,
$\rho_{\phi}$ and $\rho_{\gamma}$ are the scalar field and radiation densities,
$H ={\dot {a}}/a$ is
the Hubble parameter. 
Notice, that $\lambda$ is the brane tension which obeys the following
restriction $\lambda\geq (10TeV)^4$ \cite{Cline:1999ts, Brax:2003fv, Clifton:2011jh}. 
Obviously, substituting 
equations (\ref{1.2}),(\ref{1.22}) in Eq.(\ref{denn})
it is easy to derive the modified Klein-Gordon 
equation which describes the time evolution
of the tachyon field. This is
\begin{equation} \label{E.O.M}
\frac{\ddot{\phi}}{1-\dot{\phi}^2}+3H\dot{\phi}+\frac{V'}{V}=
-\frac{\Gamma}{V}\dot{\phi}\sqrt{1-\dot{\phi}^2} \;,
\end{equation}
where $V^{\prime}(\phi)=dV/d\phi$. 

The above cosmological equations imply that the model is 
strongly affected by the quantity $\Gamma$.
This is due to the fact that radiation is exchanging energy
with the tachyon field and this is reflected in the corresponding behavior
of dissipation coefficient $\Gamma$, which is negligible in the classical 
inflationary paradigm by definition. 
Although, the precise functional form of $\Gamma$ is still
an open issue, a number of different parametrizations
have been proposed in the literature treating the functional 
form of $\Gamma$ (see \cite{Hall:2003zp,BasteroGil:2012cm, Bastero-Gil:2014oga, Bartrum:2013fia}). In the current work 
we use the well known parametrization of 
\begin{equation}
\label{GAM}
\Gamma=\Gamma_{c}\phi^b T^c \;,
\end{equation}
where $T$ is the temperature and $\Gamma_{c}$ is constant.

During the warm inflationary epoch, the energy density of the scalar field
dominates the total fluid (stable regime \cite{BasteroGil:2012zr})
and thus Eq.(\ref{2.3}) becomes
\begin{eqnarray}
H^2=\frac{8\pi}{3M_{pl}^2}\rho_{\phi}(1+\frac{\rho_{\phi}}{2\lambda})
\end{eqnarray}
or
\begin{equation}\label{2.5}
H^{2}=\frac{8\pi}{3M_{pl}^2}\frac{V(\phi)}{\sqrt{1-\dot{\phi}^2}}
\left(1+\frac{V(\phi)}{2\lambda\sqrt{1-\dot{\phi}^2}}\right)\;.
\end{equation}
 Another important quantity in this kind of 
studies is the dimensionless dissipation parameter 
which characterizes the type of inflation 
and it is defined as (for more details see appendix)
\begin{eqnarray}\label{12}
R=\frac{\Gamma}{3H\rho_{\phi}}.
\end{eqnarray}
The above definition is presented for warm tachyon inflation in several papers \cite{Herrera:2006ck, Setare:2012fg, Setare:2013ula, Setare:2014gya, Setare:2014uja, Setare:2013dd}. Notice, that for the canonical scalar field model 
of warm inflation, the corresponding 
dimensionless ratio is defined as $\frac{\Gamma}{3H}$. 

In the weak dissipation regime, the ratio $R$ 
tends to zero 
($\Gamma/3H\rho_{\phi} \ll 1$), however, in 
the strong dissipation regime, the coefficient $\Gamma$
guides the damped evolution of the scalar field.
Now using Eq.(\ref{E.O.M}) and Eq.(\ref{2.5}) 
in the high-dissipation 
regime ($\Gamma\gg 3H\rho_{\phi}$)
to prove that 
\begin{eqnarray}\label{2.7}
\dot{\phi}^2=-\frac{3M_{pl}^2}{4\pi}\frac{(H\dot{H})}{\Gamma}
\left(1+\frac{3M_{pl}^2H^2}{4\pi \lambda}\right)^{-\frac{1}{2}} \;.
\end{eqnarray}

Owing to the fact that during inflation the parameters $H$, $\Gamma$ and $\phi$
are slowly varying functions the production of radiation become 
quasi-stable when $\dot{\rho}\ll 4H\rho_{\gamma},$ and $\dot{\rho}_{\gamma}\ll\Gamma\dot{\phi}^2$ \citep{Berera:1995ie, Berera:1996fm, Hall:2003zp}. 
Under these conditions, using Eqs.(\ref{E.O.M}) and (\ref{2.7}) we 
write the radiation density as follows:
\begin{eqnarray}\label{2.8}
\rho_{\gamma}=\frac{\Gamma \dot{\phi}^2}{4H}=-\frac{3M_{pl}^2 \dot{H}}{16\pi}
\left(1+\frac{3M_{pl}^2H^2}{4\pi\lambda}\right)^{-\frac{1}{2}}\;.
\end{eqnarray}
The latter formula can be identified with the equation relating $\rho_{\gamma}$
with the radiation temperature $T$. 
Indeed, under of adiabatic condition we may write 
\begin{equation}
\label{Kolb}
\rho_{\gamma}=C_{\gamma}T^4
\end{equation} 
where $C_{\gamma}=\frac{\pi^2g_{*}}{30}$ and $g_{*}$
is the degrees of freedom of the created massless modes \citep{Kolb1990}.
Combining Eq.(\ref{2.8}) and (\ref{Kolb}) 
we obtain the temperature
\begin{eqnarray}\label{}
T=\left(-\frac{3M_{pl}^2 \dot{H}}{16\pi C_{\gamma}}\right)^{\frac{1}{4}}
\left(1+\frac{3M_{pl}^2 H^2}{4\pi\lambda}\right)^{-\frac{1}{8}} \;.
\end{eqnarray}
Lastly, with the aid of Eqs.(\ref{2.5}) and (\ref{2.8}) we obtain the 
potential of the scalar field 
\begin{eqnarray}\label{}
& V=\lambda\left[-1+\left(1+\frac{3M_{pl}^2H^2}{4\pi\lambda}\right)^{\frac{1}{2}}\right]\\
\nonumber
&\times\left(1+\frac{3M_{pl}^2}{4\pi}\frac{(H\dot{H})}{\Gamma}
\left[1+\frac{3M_{pl}^2H^2}{4\pi \lambda}\right]^{-\frac{1}{2}}\right)^{\frac{1}{2}} \;.
\end{eqnarray}

\section{Slow-roll parameters}
Let us present here the main quantities of the tachyonic inflation. 
In particular, the basic slow-roll parameters are given by 
\begin{equation}\label{epsilon}
\epsilon=-\frac{\dot{H}}{H^2}
\end{equation}{equation}

\begin{equation}\label{epsilon1}
\eta=-\frac{\ddot{H}}{2H\dot{H}} \;.
\end{equation}{equation}
In this context, the number of e-folds is written as
\begin{equation}
N=\int_{t}^{t_{end}}H dt 
\label{efold}%
\end{equation}{equation}
where $t_{end}$ is the value of the cosmic time at the end of inflation, 
namely $\epsilon(\phi_{end})\simeq 1$ where $\phi_{end}=\phi(t_{end})$.

Also, the power spectrum of the scalar fluctuations is given by
\citep{Berera:1995ie, Berera:1996fm}
\begin{equation}
\label{PP}
P_{s}=\frac{H^2}{\dot{\phi}^2}\delta\phi^2\;.
\end{equation}{equation} 
An important feature of the warm inflationary model is related with the 
fact that the origin of $\delta \phi$ is thermal and not quantum as  
we consider in the nominal inflationary paradigm. 
In the case of warm inflation 
it has been found \citep{Berera:1995ie, Berera:1996fm, Hall:2003zp}
that scalar perturbations are written as  
\begin{eqnarray}\label{2.12}
\delta\phi^2\simeq\frac{k_F T}{2 M_{pl}^4\pi^2} \;,
\end{eqnarray}
where the wave number 
$k_F=\sqrt{\frac{\Gamma H}{V}}=H\sqrt{\frac{\Gamma}{3HV}}\geq H$ 
corresponds to the freeze-out scale at the special point when, 
the dissipation damps out to thermally excited 
fluctuations of inflaton 
($\frac{V''}{V'}<\frac{\Gamma H}{V}$) \cite{Taylor:2000ze}. 
 Notice, that Eq.(\ref{2.12}) is valid in the 
high-dissipation regime $R\gg 1$. 
As we have already mentioned in the previous section 
we study our model via Eq.(\ref{2.7}) in the context of 
high-dissipation regime, which means that  
for scalar perturbations we can utilize Eq.(\ref{2.12}).
 
Inserting the freeze-out wave-number and Eq.(\ref{2.12}) into 
Eq.(\ref{PP}) we find after some simple calculations that the power spectrum of the tachyonic scalar field is given by
\begin{eqnarray}\label{scalar power spectrum}
P_{s}\simeq \frac{H^{\frac{5}{2}}\Gamma^{\frac{1}{2}} T}{2\pi^2 M_{pl}^4 V^{\frac{1}{2}}\dot{\phi}^2} \;.
\end{eqnarray}
 We would like to point out that in the case of 
canonical scalar fields within the framework of warm inflation 
one can find other forms of the power spectrum $P_{s}$ 
in Refs.\citep{Bartrum:2013fia,Ramos:2013nsa,Bastero-Gil:2014jsa}.
The corresponding spectral index $n_{s}$ is defined in terms of the
$P_{s}$ slow-roll parameters, as usual \cite{LL}, by
\begin{eqnarray}\label{spectral index}
n_s-1=\frac{d\ln P_{s}}{d\ln k} \;.
\end{eqnarray}

On the other hand, it has been found \cite{Langlois:2000ns} that the 
power spectrum of the tensor perturbations which are defined on the brane 
takes the form
\begin{eqnarray}\label{tensor power spectra}
P_t=\frac{64\pi}{M_{pl}^2}\left(\frac{H}{2\pi}\right)^2
G^2(x) 
\end{eqnarray}
where $x\equiv \left[\frac{3HM_{pl}^2}{4\pi\lambda}\right]^{\frac{1}{2}}$
and $G(x)=[\sqrt{1+x^2}-x^2\sinh^{-1}(\frac{1}{x})]^{-\frac{1}{2}}$ arises 
from normalization of zero-mod of a graviton \cite{Langlois:2000ns}. 
Therefore, using the so called tensor-to-scalar ratio 
we arrive at
\begin{eqnarray}\label{Tensor to scalar ratio}
r=32\pi M_{pl}^2\frac{V^{\frac{1}{2}}\dot{\phi}^2}{H^{\frac{1}{2}}\Gamma^{\frac{1}{2}} T}G^2(x)\;.
\end{eqnarray}

In order to proceed with the analysis it would help to know the functional 
form of the scale factor $a(t)$. 
Barrow \cite{Barrow:1996bd} showed that under of specific conditions 
we can have an intermediate inflation in which the scale factor 
satisfies the following exponential form:
\begin{eqnarray}\label{scall}
a(t)=a_{I}\exp(At^f)\;, 
\end{eqnarray}
which provides
\begin{eqnarray}\label{scal}
H(t)=\frac{{\dot a}(t)}{a(t)}=Aft^{f-1}\;,
\end{eqnarray}
where $f$ satisfies the restriction $0<f<1$.
The above expansion evolves faster than the power-law inflation
($a \propto t^{p}$, $p>1$)
and slower than the standard deSitter one, $a(t)\propto {\rm exp}(H_{I}t)$
[$H(t)=H_{I}=$const.].
 Considering the functional form (\ref{GAM}), 
one has to deal in general with the following
four parametrizations, which have been considered
within different approaches in the literature.
Depending on the values of $(b,c)$ we have: 
(I)- the situation in which the formula is $\Gamma=\Gamma_3 T^3\phi^{-2}$, 
$(b,c)=(3,-2)$. 
The constant parameter $\Gamma_3$ corresponds to 
$0.02 h^2 \mathcal{N}_{Y}$ where  
there is generic supersymmetric (SUSY) model 
with chiral superfields $X$, $\Phi$ and $Y_i=1,...\mathcal{N}_{Y}$. 
This case is mostly used in the low temperature regime 
where $m_{\chi}$ ($m_{\chi} $ is the mass of 
catalyst field) \citep{Bastero-Gil:2014oga, Bastero-Gil:2013nja}; 
(II)- for $(b,c)=(2,-1)$ we have 
$\Gamma=\Gamma_2\phi^2 T^{-1}$. This parametrization has been used 
for non-SUSY models \cite{Berera:1998gx, Yokoyama:1998ju}.
Q5: (III)- the case where 
$\Gamma=\Gamma_{0}$ is a positive 
constant (hereafter $\Gamma_{0}$-parametrization:
see \cite{Herrera:2006ck, Deshamukhya:2009wc, Setare:2012fg, Setare:2013ula, Setare:2013qfa, Setare:2014gya, Setare:2014uja, Setare:2015cta, Setare:2013dd})  
which implies that the pair $(b,c)$ in Eq.(\ref{GAM}) 
is strictly equal to $(0,0)$
and finally (IV) 
we utilize the so called high temperature regime 
(hereafter $\Gamma_{1}$-parametrization) 
in which we select  
$(b,c)=(0,1)$ and thus $\Gamma \propto  T$
(see also \cite{Panotopoulos:2015qwa}). 
In this paper, we are going to focus
on parameterizations (III) and (IV) in order 
to calculate the slow roll parameters. 
Lastly, we remind the reader that in the framework of warm inflationary
model thermal fluctuations dominate over the quantum fluctuations.


  Combining the latter argument with the fact that 
thermal fluctuations are proportional to temperature $T$ while 
quantum fluctuations are proportional to $H$, one can easily derive the 
condition $T>H$. Obviously,  
if we consider our model in the high temperature regime ($\Gamma \propto T$)
then the aforesaid restriction ($T>H$) is satisfied.
For more  details we refer the reader the work of \cite{Panotopoulos:2015qwa}. 
 



\subsection{$\Gamma_{0}$-parametrization }\label{}
In this inflationary scenario ($\Gamma=\Gamma_{0}$=const.)
with the aid of Eq.(\ref{scal}) we integrate
Eq.(\ref{2.7}) and we obtain  
the evolution of the scalar field in terms of the 
hyper-geometric function \cite{Arfken,Abramowitz} 
\begin{eqnarray}\label{sola}
\phi(t)-\phi_0=\frac{F(t)}{K}
\end{eqnarray}
where 
\begin{eqnarray}\label{}
F(t)&=&t^{\frac{2f-1}{2}} \times \\ \nonumber &_{2}F_1&\left[\frac{1}{4},\frac{1-2f}{4(1-f)},\frac{5-6f}{4(1-f)},-\frac{3M_{pl}^2f^2A^2t^{2f-2}}{4\pi\lambda }\right] \;,
\end{eqnarray} 
$$K=-\left(\frac{16\pi\Gamma_0(1-f)}{3M_{pl}^2f^2A^2}\right)^{\frac{1}{2}}\frac{\Gamma(\frac{5-6f}{4(1-f)})}{\Gamma(\frac{1-2f}{4(1-f)})}$$ 
and $\Gamma(n)$ is the normal 
Gamma-function. Notice, that 
without losing the generality we have set
$\phi_0=0$.

Now we can derive the Hubble parameter $H$ and the 
associated potential $V(\phi)$ 
in the limit of $\dot{\phi}^2\ll V(\phi)$ 
\begin{eqnarray}\label{HHA}
&H(\phi)=fA(F^{-1}[K\phi])^{f-1}\\
\nonumber
&V(\phi)\simeq\lambda(-1+\sqrt{1+
\frac{3M_{pl}^2f^2A^2[F^{-1}(K\phi)]^{2f-2}}{4\pi\lambda}})
\end{eqnarray}
where $F^{-1}(\phi)$ is the inverse function of $F(t)$. 
Clearly, if we substitute Eq.(\ref{HHA}) in the slow-roll parameters 
then we have 
\begin{equation}
\label{EEPS}
\epsilon=\frac{(1-f)t^{-f}}{fA}=
\frac{1-f}{fA[F^{-1}(K\phi)]^f}
\end{equation}
\begin{equation} 
\label{EETA}
\eta=\frac{(2-f)t^{-f}}{2fA}=
\frac{2-f}{2fA[F^{-1}(K\phi)]^f} \;.
\end{equation} 
Notice, that in order to extract the latter equalities in 
Eqs(\ref{EEPS}), (\ref{EETA}) we used Eqs.(\ref{scal}), (\ref{sola}).
In our warm intermediate case the condition 
$\epsilon=1$, insures the beginning of inflation 
\cite{Barrow:2006dh, Barrow:1993zq}. 
Therefore, utilizing 
Eq.(\ref{efold}), 
we can derive the number of e-folds\footnote{In the
literature sometimes we replace $\phi$ by $\phi_{\star}$ which denotes the value
at the horizon crossing.}
 \begin{eqnarray}\label{number of efolds}
&N=\int_{t_{in}}^{t_{*}} Hdt=A(t_{*}^f-t_{in}^f)&\\
\nonumber
&=A\left( [F^{-1}(K\phi)]^f-[F^{-1}(K\phi_{in})]^f\right)& \;.
\end{eqnarray}
Plugging $\phi_{in}$ into Eq.(\ref{EEPS}) and using the 
constraint $\epsilon(\phi_{in})\simeq 1$ we find 
\begin{equation}
\label{ee1}
\phi_{in}=\frac{1}{K}F(y),\;\;\;\;\;
y=\left(\frac{1-f}{fA}\right)^{\frac{1}{f}} \;. 
\end{equation}
In order to proceed with the analysis we need to know the values of $N$ and
$\phi_{in}$. 
Firstly, it is natural to consider that the number of e-folds
is 50 or 60.
Secondly, using the condition $\epsilon(\phi_{in})=1$ and
Eqs. (\ref{number of efolds}), (\ref{ee1}) 
we can estimate the slow-roll parameters.


Now we focus on the power spectrum formulas. 
Specifically, inserting the appropriate expressions into 
Eq.(\ref{scalar power spectrum}) we define 
the scalar power spectrum 
\begin{eqnarray}\label{}
&P_{s}=p_1I(N)^{\frac{3f}{4}}\left[-1+\left(1+\frac{3M_{pl}^2f^2A^2}
{4\pi\lambda I(N)^{2-2f}}\right)^{\frac{1}{2}}\right]^{-\frac{1}{2}}\\
\nonumber
&\times \left(1+\frac{3M_{pl}^2f^2A^2}{4\pi\lambda  I(N)^{2-2f}}\right)^{\frac{3}{8}}
\end{eqnarray}
where $p_1=(\frac{\Gamma_0^6 f^3A^3}{3^3M_{pl}^6\pi^5(1-f)^3C_{\gamma}})^{\frac{1}{4}}$ and $I(N)=\left[\frac{1+f(N-1)}{fA}\right]^{\frac{1}{f}}$.
Combining the definition of the spectral index $n_{s}$ 
(\ref{spectral index}) and the above equation we obtain 
\begin{eqnarray}\label{ns-gama0}
n_s-1=-\frac{3}{4A}I(N)^{-f}+n_1+n_2
\end{eqnarray}
where in the derivation of the above equality we have used
\begin{eqnarray}\label{}
&n_1=\frac{3M_{pl}^4(f-1)fA}{8\pi\lambda} I(N)^{f-2}
\left(1+\frac{3M_{pl}^2f^2A^2}{4\pi\lambda I(N)^{2-2f}}\right)^{-\frac{1}{2}}\\
&\nonumber
\times \left[-1+(1+\frac{3M_{pl}^2f^2A^2}{4\pi\lambda I(N)^{2-2f}})^{\frac{1}{2}}\right]^{-1}\\
\nonumber
&n_2=\frac{9M_{pl}^2(1-f)fA}{16\pi\lambda} I(N)^{f-2}
\left(1+\frac{3M_{pl}^2f^2A^2}{4\pi\lambda  I(N)^{2-2f}}\right)^{-1} \;.
\end{eqnarray}
Lastly, based on Eq.(\ref{Tensor to scalar ratio}) we compute 
the tensor-to-scalar ratio parameter 
\begin{eqnarray}\label{r-gama0}
r=r_1I(N)^{\frac{5f-8}{4}}\left(1+\frac{3M_{pl}^2f^2A^2}{4\pi\lambda I(N)^{2-2f}}\right)^{-\frac{3}{8}}\\
\nonumber
\left[-1+\left(1+\frac{3M_{pl}^2f^2A^2}{4\pi\lambda  I(N)^{2-2f}}\right)^{\frac{1}{2}}\right]^{\frac{1}{2}}G^2(N)
\end{eqnarray}
where $r_1=(\frac{2^{16} \lambda^2 3^3 C_{\gamma} \pi (fA)^5 (1-f)^3}{M_{pl}^2\Gamma_0^6})^{\frac{1}{4}}$. 


\subsection{$\Gamma_{1}$-parametrization}\label{}
Using the same methodology as in the previous section  
we provide the basic slow-roll parameters in the case of
$\Gamma_{1}$-parametrization, namely $\Gamma=\Gamma_1T$, where $\Gamma_{1}$
is constant. In particular, from Eqs.(\ref{scal},\ref{2.7})
the tachyon field is written as 
\begin{eqnarray}\label{PPL}
\phi-\phi_0=\frac{{\tilde F}(t)}{\tilde{K}}
\end{eqnarray}
where ${\tilde F}(t)$ and $\tilde{K}$ are given by the following
expressions
\begin{eqnarray}\label{}
&{\tilde F}(t)=t^{\frac{7f-2}{8}}\times \\ 
\nonumber 
 &_{2}F_1\left[\frac{3}{16},\frac{7f-2}{16(f-1)},
\frac{23f-18}{16(f-1)},-\frac{3M_{pl}^2f^2A^2}{4\pi\lambda}t^{2f-2}\right]\\
\nonumber
&\tilde{K}=-\left(\frac{2^6\pi\Gamma_1^4 (1-f)^5}{3M_{pl}^2C_{\gamma}f^7A^7}\right)^{\frac{1}{8}}\frac{\Gamma(\frac{23f-18}{16(f-1)})}{\Gamma(\frac{7f-2}{16(f-1)})}\;.
\end{eqnarray}
Also here we have set $\phi_0=0$. 

Now the Hubble parameter, the potential and the corresponding slow-roll
parameters are given by
\begin{eqnarray}\label{HUBB}
&H(\phi)=fA({\tilde F}^{-1}[{\tilde K}\phi])^{f-1}\\
\nonumber
&V(\phi)\simeq\lambda(-1+\sqrt{1+
\frac{3M_{pl}^2f^2A^2[{\tilde F}^{-1}({\tilde K}\phi)]^{2f-2}}{4\pi\lambda}})
\end{eqnarray}
\begin{equation} 
\epsilon=\frac{1-f}{fA[{\tilde F}^{-1}({\tilde K}\phi)]^f} \;,
\end{equation}
\begin{equation} 
\eta=
\frac{2-f}{2fA[{\tilde F}^{-1}({\tilde K}\phi)]^f} \;.
\end{equation} 
where ${\tilde F}^{-1}(K\phi)$ is inverse function of ${\tilde F}(t)$. 
In the current case the 
number of e-folds becomes 
 \begin{eqnarray}\label{number of efolds2}
&N=\int_{t_{in}}^{t_{*}} Hdt=A(t_{*}^f-t_{in}^f)&\\
\nonumber
&=A\left( [{\tilde F}^{-1}({\tilde K}\phi)]^f-[{\tilde F}^{-1}({\tilde K}\phi_{in})]^f\right)& \;.
\end{eqnarray}
and following standard lines
the end of inflation takes place when 
\begin{equation}
\phi_{in}=\frac{1}{\tilde{K}}\tilde{F}(y),\;\;\;\;\;\;
y=\left(\frac{1-f}{fA}\right)^{\frac{1}{f}} \;. 
\end{equation}

The scalar power-spectrum can be easily identified by 
comparing the current cosmological expressions with 
Eq.(\ref{scalar power spectrum}), and we find
\begin{eqnarray}\label{}
&P_{s}=p_2 I(N)^{\frac{9f-6}{8f}}\left[-1+\left(1+\frac{3M_{pl}^2f^2A^2}{4\pi\lambda I(N)^{2-2f}}\right)^{\frac{1}{2}}\right]^{-\frac{1}{2}}\\
\nonumber
&\times\left[1+\frac{3M_{pl}^2f^2A^2}{4\pi\lambda I(N)^{2-2f}}\right]^{\frac{3}{16}}
\end{eqnarray}
where 
$p_2=\frac{2(\Gamma_1 fA)^{\frac{3}{2}}}{3M_{pl}^2\lambda^{\frac{1}{2}}}(\frac{16\pi C_{\gamma}}{3M_{pl}^2(1-f)})^{\frac{3}{8}}$. 
If we take the aforementioned $P_{s}$ formula we find 
the following spectral index 
\begin{eqnarray}\label{ns-gama1}
n_s-1=-\frac{9f-6}{8fA} I(N)^{-f}+n_1+n_2
\end{eqnarray}
where
\begin{eqnarray}\label{}
&n_1=\frac{3M_{pl}^4(f-1)fA}{8\pi\lambda} I(N)^{f-2}
\left(1+\frac{3M_{pl}^2f^2A^2}{4\pi\lambda  I(N)^{2-2f}}\right)^{-\frac{1}{2}}\\
&\nonumber
\times \left[-1+(1+\frac{3M_{pl}^2f^2A^2}{4\pi\lambda I(N)^{2-2f}})^{\frac{1}{2}}\right]^{-1}\\
\nonumber
&n_2=\frac{9M_{pl}^2(1-f)fA}{32\pi\lambda} I(N)^{f-2}\left(1+\frac{3M_{pl}^2f^2A^2}{4\pi\lambda I(N)^{2-2f}}\right)^{-1}
\end{eqnarray}
Finally, the tensor-to-scalar ratio [see Eq.{\ref{Tensor to scalar ratio})]
takes the form 
\begin{eqnarray}\label{r-gama1}
r=r_2 I(N)^{\frac{7f-6}{8}}\left(1+\frac{3M_{pl}^2f^2A^2}{4\pi\lambda I(N)^{2-2f}}\right)^{\frac{-3}{16}}\\
\nonumber
\left[-1+(1+\frac{3M_{pl}^2f^2A^2}{4\pi\lambda I(N)^{2-2f}})^{\frac{1}{2}}\right]^{\frac{1}{2}}G^2(N)
\end{eqnarray}
where $r_2=\frac{24\lambda^{\frac{1}{2}}(fA)^{\frac{7}{8}}}{\Gamma_1^{\frac{3}{2}}}
\left(\frac{16\pi C_{\gamma}(1-f)}{3M_{pl}^2}\right)^{\frac{3}{8}}$.

\section{Comparison with observation}\label{observation}
The analysis of \textit{Planck} \cite{Ade:2015lrj} and 
BICEP2/Keck Array \cite{joint} data sets has provided 
a new constraint on inflationary scenarios~\cite{Martin:2013tda}. 
In particular, the comprehensive analysis 
of \textit{Planck} data~\cite{Ade:2015lrj} 
indicates that single 
scalar-field models of slow-roll inflation have a very low tensor-to-scalar
fluctuation ratio $r=P_{t}/P_{s}\ll1$, a scalar spectral index
$n_{s}=0.968\pm 0.006$ and no appreciable running. 
The upper bound set by 
the \textit{Planck} team and the joint analysis of
BICEP2/Keck Array/\textit{Planck} \cite{joint}
on tensor-to-scalar fluctuation ratio is $r<0.11$.
In this section we attempt to test the performance of the warm inflationary
model against the above observational results.

Let us now concentrate on our results. 
Notice, that in the case of warm inflation 
the number of degrees of freedom becomes 
$g_{*} \simeq 200$ ($C_{\gamma}\simeq 70$,\cite{BasteroGil:2006vr})
Also, for the rest of the paper 
we have set $\lambda=10^{-14}$. Concerning 
the number of e-folds, it is natural to consider that $N$ 
lies in the interval $[50,60]$. Here, we have set it either to 
50 or 60.
In figures 1 ($\Gamma_{0}$ parametrization) 
and 2 ($\Gamma_{1}$ parametrization) 
we present the $A-f$ allowed region 
in which our $(n_{s},r)$ results satisfy the above restrictions of 
{\em Planck} within 1$\sigma$ uncertainties.
In the case of $\Gamma_0$ model, we observe that for various values
of the dissipation coefficient there is a narrow region in the $A-f$ plane 
which is consistent with the observed values of $n_{s}$ and $r$.
The absence of $A-f$ pair solutions and thus of $(n_{s},r)$, appear for 
$\Gamma_0\le 10^{-10}$.

For the $\Gamma_1$ parametrization the situation is slightly 
different. 
Figure (\ref{fig-gama1}) shows 
broader $A-f$ regions with respect to those 
of $\Gamma_{0}$ parametrization.
Also in this case we verify 
that for $\Gamma_1\le 10^{-10}$, there is no $A-f$ pairs 
which satisfy the observational criteria.
 The theoretical curves of cold intermediate 
inflation model in Einstein General Relativity in $n_s-r$ plane are well outside of the $95 \%$
C.L region. 
Our aim here is to test the viability of the warm inflation, involving the 
latest {\it Planck2015} data. In figure (\ref{fig-planck}) we present the
confidence contours in the $(n_{s},r)$ plane. 
On top of figure (\ref{fig-planck}) we provide 
the solid stars for the individual 
sets of $(n_{s},r)$ which are based on the
$\Gamma_{0}$ parametrization, whereas in the same figure we 
display the corresponding solid points 
in the case of $\Gamma_{1}$ parametrization. 
From the comparison it becomes clear that our 
$(n_{s},r)$ results are in excellent agreement 
with those of \textit{Planck 2015}. 

Indeed, we find:

(a) $\Gamma_{0}$ parametrization: if we use $N=50$ then we find 
$n_{s}=0.9675$ and $r=0.0086$, whereas for $N=60$ we have  
$n_{s}=0.9627$ and $r=0.0036$. 

(b) $\Gamma_{1}$ parametrization: in the case of $N=50$ we obtain
$n_{s}=0.9638$ and $r=0.00857$ and for $N=60$ we have  $n_{s}=0.9692$ and 
$0.00187$.  

Bellow we compare the current predictions with those 
of viable literature potentials. 
This can help us to understand the variants  
of the warm inflationary model
from the observationally viable inflationary scenarios.

\begin{itemize}

\item The chaotic inflation \cite{Linde}:
In this inflationary model the potential is 
$V(\phi) \propto \phi^{k}$. 
Therefore, the basic slow-roll
parameters are written as $\epsilon=k/4N$, $\eta=(k-1)/2N$
which implies $n_{s}=1-(k+2)/2N$ and $r=4k/N$. 
It has been found that 
monomial potentials with $k\ge 2$ can not accommodate the Planck
priors \cite{Ade:2015lrj}. 
For example, using $k=2$ and $N=50$ we obtain 
$n_{s}\simeq 0.96$ and $r\simeq 0.16$. For 
$N=60$ we have $n_{s}\simeq 0.967$ and $r\simeq 0.133$. 
It is interesting to mention that 
the chaotic inflation also
corresponds to the slow-roll regime of intermediate inflation \cite{Barrow:1990vx,Barrow:1993zq,Barrow:2006dh,Barrow:2014fsa} with
Hubble rate during inflation given by $H\propto t^{k/(4-k)}$ with
$n_{s}=1-(k+2)r/8k$ and $k=-2$ gives $n_{s}=1$ exactly to first order. 

\item The $R^{2}$ inflation \cite{staro}:
In Starobinsky inflation the asymptotic behavior 
of the effective potential becomes 
$V(\phi)\propto\lbrack1-2\mathrm{e}%
^{-B\phi/M_{pl}}+\mathcal{O}(\mathrm{e}^{-2B\phi/M_{pl}})]$
which provides the following slow-roll predictions \cite{Muk81,Ellis13}:
$n_{s}\approx 1-2/N$ and $r\approx 8/B^{2}N^{2}$, where $B^{2}=2/3$.
Therefore, if we select $N=50$ then we obtain
$(n_{s},r)\approx(0.96,0.0048)$. For $N=60$ we
find $(n_{s},r)\approx(0.967,0.0033)$. 
It has been found that the Planck data \cite{Ade:2015lrj}
favors the Starobinsky inflation. Obviously, our results 
(see figure 3) are consistent with those of $R^{2}$ inflation.

\item Hyperbolic inflation \cite{Basilakos:2015sza}:
In hyperbolic inflation the potential is
given by $V(\phi) \propto \mathrm{sinh}^{b}(\phi/f_{1})$.
Initially, Rubano and Barrow  \cite{Rubano:2001xi}
proposed this potential in the context of dark energy.
Recently, Basilakos \& Barrow \cite{Basilakos:2015sza} investigated the properties
of this scalar field potential back in the inflationary epoch.
Specifically, the slow-roll parameters are written as
$$
\epsilon=\frac{b^{2}M_{pl}^{2}}{2f_{1}^{2}}\mathrm{coth}^{2}(\phi/f_1), \label{ee1}%
$$
$$
\eta=\frac{bM_{pl}^{2}}{f_{1}^{2}}\left[  (b-1)\mathrm{coth}^{2}(\phi/f_1)+1\right]
$$
and 
$$
\phi=f_1\;\mathrm{cosh}^{-1}\left[  e^{NbM_{pl}^{2}/f^{2}}\mathrm{cosh}%
(\phi_{end}/f_1)\right]  . \label{efold2}%
$$
where $\phi_{end}\simeq\frac{f}{2}\mathrm{ln}\left(  \frac{\theta+1}{\theta
-1}\right)$. Comparing this model with the data 
Basilakos \& Barrow \cite{Basilakos:2015sza} found 
$n_{s}\simeq 0.968$, $r\simeq 0.075$, $1<b \le 1.5$ 
and $f_1\ge 11.7M_{pl}$.

\item Other inflationary models:
The origin of brane \cite{Dvali:2001fw,GarciaBellido:2001ky} and exponential \cite{Goncharov:1985yu,Dvali:1998pa} 
inflationary models are motivated by the physics of 
extra dimentions and supergravity respectively. 
It has been found that these models are in agreement
with the Planck data 
although the $R^{2}$ inflation 
is the winner from the comparison \cite{Ade:2015lrj}.

\end{itemize}


\begin{figure}
\begin{center}
\includegraphics[scale=0.47]{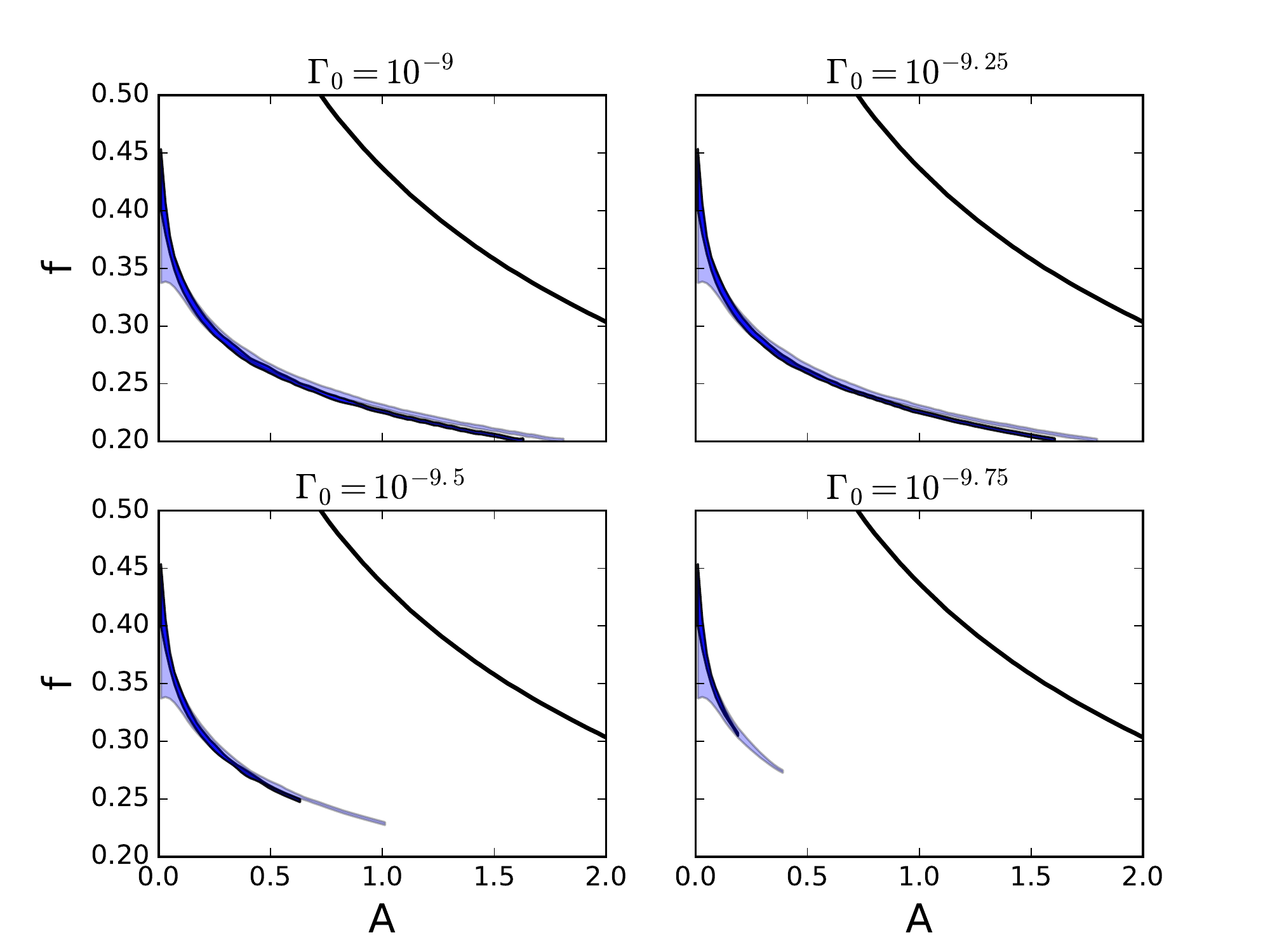}
\end{center}
\caption{The $A-f$ diagram which coincides 
within 1$\sigma$ confidence level of {\em{Planck}} data. The corresponding 
values of 
$\Gamma_0$ are shown at the top of panels. 
The background transparent (foreground opaque) indicates $N=60$ 
($N=50$). These values of $\Gamma_0$ are consistent with $R\gg1$. The solid black curve shows the boundary $T=H$ and below the curve is consistent with $T>H$.}
\label{fig-gama0}
\end{figure}

\begin{figure}
\begin{center}
\includegraphics[scale=0.47]{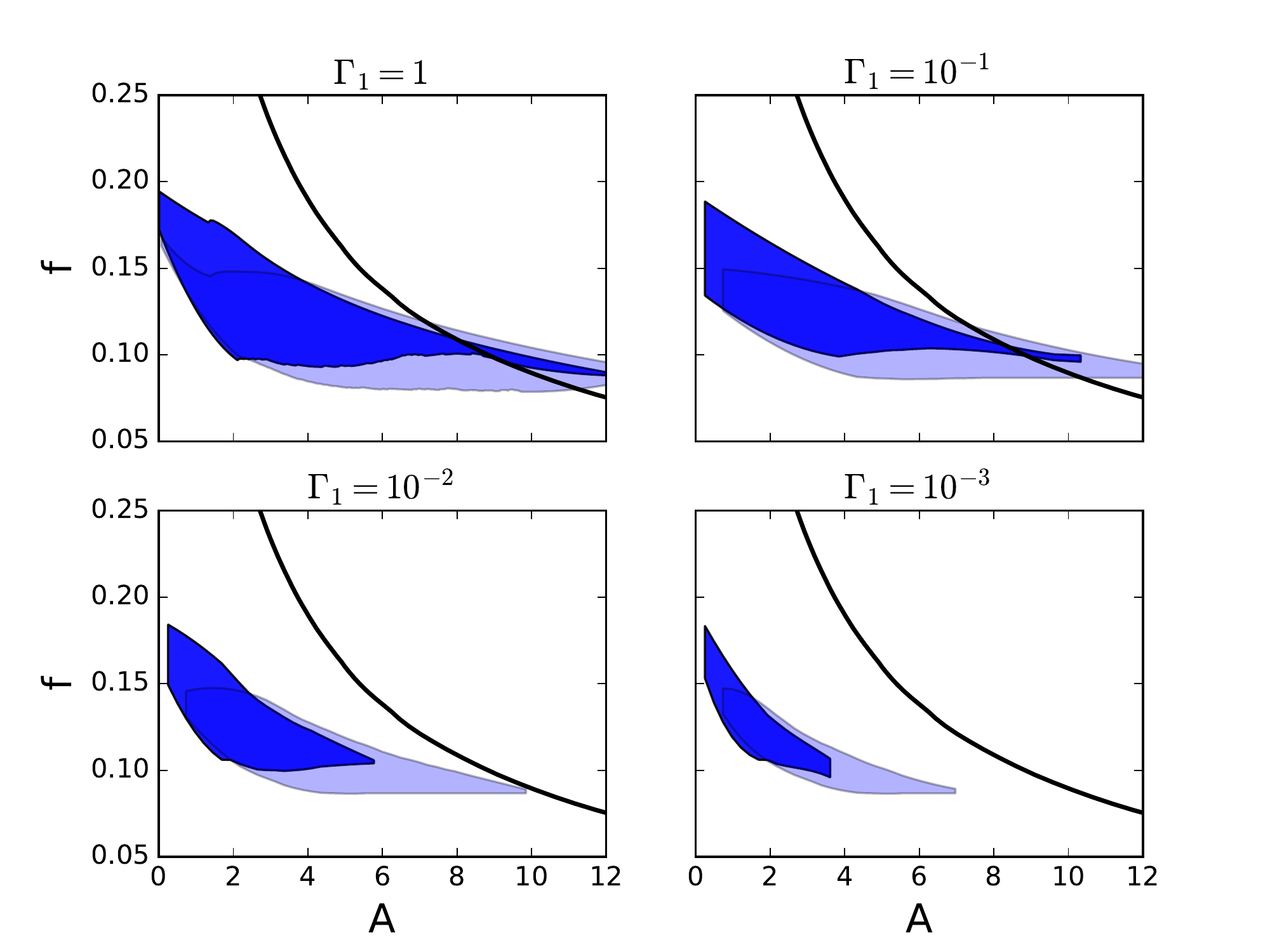}
\end{center}
\caption{The $A-f$ region in the case of $\Gamma_1$ parametrization. These values of $\Gamma_1$ are consistent with $R\gg1$. The solid black curve is same as Fig.(\ref{fig-gama0}). }
\label{fig-gama1}
\end{figure}

\begin{figure}
\begin{center}
\includegraphics[scale=0.47]{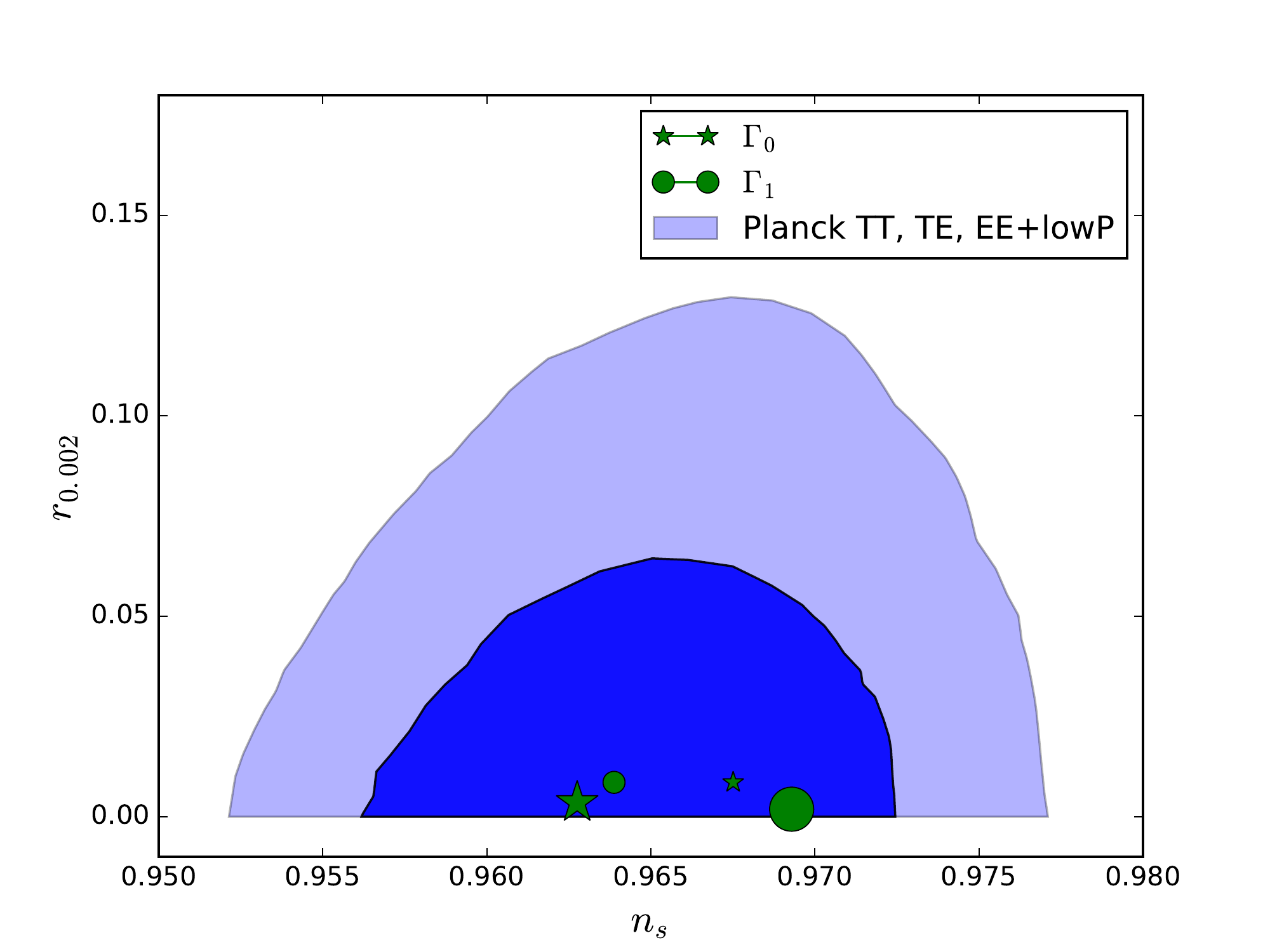}
\end{center}
\caption{1$\sigma$ and 2 $\sigma$ confidence regions borrowed from 
{\em{Planck}} \cite{Ade:2015lrj}. 
Stars (squares) indicate warm inflation with 
$\Gamma_0$ ($\Gamma_1$) parametrization. Big and small points correspond to 
$N=60$ and $N=50$ respectively.  For $\Gamma_0$ we set $f=0.28$, $A=0.35$ and $\Gamma_0=10^{-9}$. For $\Gamma_1$ we set $f=0.13$, $A=3.$ and $\Gamma_1=10^{-3}$. 
}
\label{fig-planck}
\end{figure}

 At this point we would like to mention that 
in the high-dissipation regime $R\gg 1$, there is always a region in $A-f$ 
plane which is consistent with the warm inflation condition $T>H$. 
To clarify this issue we plot in Fig.(\ref{fig:warm}) 
the diagram of $\log_{10}\frac{T}{H}$ in the $A-f$ plane.
The solid line corresponds to the boundary limit $T=H$. 
Clearly, based on the condition $T>H$
we can reduce the 
parameter space and thus producing one of the strongest 
existing constraints (to our knowledge) on $A$ and $f$.
Note that in order to produce the above diagram we have 
fixed the initial values of $T$ and $H$ to those at the beginning of inflation. 
After the triggering of inflation the
inflaton/photon interaction takes place which leads to radiation production and thus it guarantees that the above condition holds during the inflationary era.
\begin{figure}
	\begin{center}
		\includegraphics[scale=0.47]{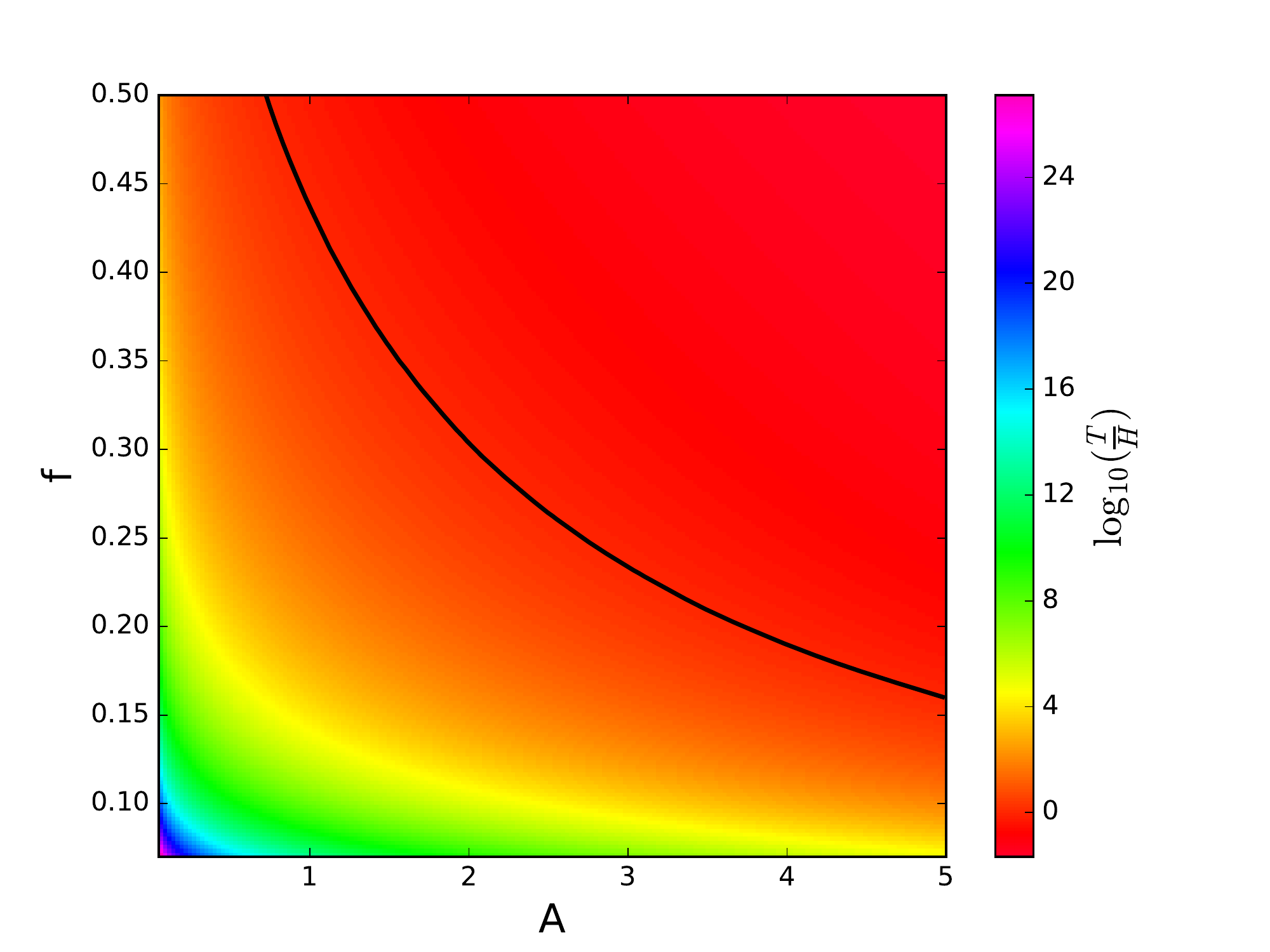}
	\end{center}
	\caption{The value of $\frac{T}{H}$ at beginning of inflation in A-f plane. The solid black curve shows boundary $T=H$.}
	\label{fig:warm}
\end{figure}

 Finally, we investigate the possibility to treat 
$\lambda$ as a free parameter. 
In fact there are three main conditions which we need to use in order 
to provide a viable limit on the $\lambda$. These are: (a) the 
high dissipation regime $R\gg 1$, (b) the warm inflation condition $T>H$ 
and (c) to recover the {\it Planck2015} $(n_s,r)$ observational constraints. 
Our investigation shows that $\lambda$ is correlated 
with the $(A,f)$ pair. For example for $(A,f)=(0.5,0.25)$ we find 
$\lambda>6\times10^{-16}$ which is consistent with above conditions 
while for $(A,f)=(0.4,0.2)$ we obtain $\lambda>1.5\times10^{-20}$. 
In general we verify that it is not possible 
to find a lower value of $\lambda$ for all pairs of $(A,f)$.

\section{Appendix}\label{Appendix}
In this paper we have studied our model in natural unit ($\frac{h}{2\pi}=c=1$)
	 therefore we have ($[\rm mass]=M$, $[\rm time]=T$ and $[\rm length]=L$ where $[A]$ means dimension of "$A$")
\begin{eqnarray}\label{}
[c]=LT^{-1}=1~~~~~~~~~~~~~[h]=M L^2 T^{-1}\\
\nonumber
\Rightarrow~~~~T=L=M^{-1}~~~~~~~~~~~~~~~~.
\end{eqnarray}
Using Eq.(\ref{2.3}) we have
\begin{eqnarray}\label{}
[H^2]=[\frac{8\pi}{M^2}\rho_{\phi}(1+\frac{\rho_{\phi}}{2\lambda})]~~~~~~~~~~~~~~~~~~~~~~~\\
\nonumber
\Rightarrow \frac{[a^2]}{[a^2] T^2}=\frac{[\rho_{\phi}]}{[M^2]}\Rightarrow~~~[\rho_{\phi}]=[T_{\mu}^{\nu}]=[P_{s}]=M^4
\end{eqnarray}
where $\rho_{\phi}$ is the scalar field energy density with dimension $M^4$.
From Eq.(\ref{1.2}) we have 
\begin{eqnarray}\label{}
[\dot{\phi}]=1~~~\Rightarrow~~~~[\phi]=M^{-1}
\end{eqnarray}
It appears that the tachyon scalar field has dimensions of $M^{-1}$. In Eq.(\ref{denn}) r.h.s and l.h.s have dimension $M^4$
\begin{eqnarray}\label{}
[\dot{\rho_{\phi}}]+[3H\rho_{\phi}]+[3HP_{\phi}]=[\Gamma \dot{\phi}^2]\\
\nonumber
\Rightarrow~~\frac{[\rho_{\phi}]}{T}+\frac{[\rho_{\phi}]}{T}+\frac{[P_{\phi}]}{T}=[\Gamma]\\
\nonumber ~~\Rightarrow [\Gamma]=M^{5}~~~~~~~~~~~~~~~.
\end{eqnarray} 
Now based on 
Eq.(\ref{12}) we find 
\begin{eqnarray}\label{}
[R]=\frac{[\Gamma]}{[H][\rho_{\phi}]}=\frac{M^5}{M M^4}=1 .
\end{eqnarray} 


\section{Conclusions}\label{conclusion}
In this article we investigate 
the warm inflation for the
Friedmann-Robertson-Walker spatially
flat cosmological model in which the 
scale factor of the universe satisfies the form of 
Barrow \cite{Barrow:1996bd}, namely  
$a(t)=a_I\exp(At^f)$ ($ 0<f<1$). 
Within this context, we estimate analytically the 
slow-roll parameters and we compare our predictions 
with those of other inflationary models as well as
we test the performance of warm inflation against
the observational data. 
We find that currently warm 
inflationary model 
is consistent with the results given by \emph{Planck 2015} within
$1\sigma$ uncertainties.



\begin{acknowledgements}
SB acknowledges support by the Research Center for
Astronomy of the Academy of Athens in the context of the program
\textquotedblleft\textit{Tracing the Cosmic Acceleration}.
\end{acknowledgements}

 \bibliographystyle{spphys}
  \bibliography{ref}

\end{document}